# Design-to-Robotic-Production of Underground Habitats on Mars

H. Bier[1,2,3], H. Vermeer[1], A. Hidding[1], and K. Jani[1]

[1] TU Delft
[2] Politecnico Milano
[3] UAS Anhalt

Keywords
Data-driven design, robotic production and operation, habitat, renewable energy, autarkic system

Abstract
In order for off-Earth top surface structures built from regolith to protect astronauts from radiation, they need to be several meters thick. Technical University Delft (TUD) proposes to excavate into the ground to create subsurface habitats. By excavating not only natural protection from radiation can be achieved but also thermal insulation because the temperature is more stable underground. At the same time through excavation valuable resources can be mined for through in situ resource utilization (ISRU). The idea is that a swarm of autonomous mobile robots excavate the ground in a sloped downwards spiral movement. The excavated regolith will be mixed with cement, which can be reproduced on Mars through ISRU, in order to create concrete. The concrete is 3D printed/sprayed on the excavated tunnel to reinforce it. As soon as the tunnels are reinforced, the material in-between the tunnels can be removed in order to create a larger cavity that can be used for inhabitation. Proposed approach relies on Design-to-Robotic-Production (D2RP) technology developed at TUD[1] for on-Earth applications. The rhizomatic 3D printed structure is a structurally optimized porous shell structure with increased insulation properties. In order to regulate the indoor pressurised environment an inflatable structure is placed in the 3D printed cavity. This inflatable structure is made of materials, which can also be at some point reproduced on Mars through ISRU. Depending on location the habitat and the production system are powered by a system combining solar and kite-power[2]. The ultimate goal is to develop an autarkic D2RP system for building subsurface autarkic habitats on Mars from locally obtained materials.

Introduction
Building habitats on Mars requires the acknowledgment of three interconnected aspects: First, the Design-to-Robotic-Production (D2RP) methodology developed at Technical University Delft (TUD) for on-Earth applications has to be adapted to Mars conditions, second, the geology and available materials, the climate and possible hazards and their impact on the D2RP process have to be considered, and third, the limits in terms of mass and volume for interplanetary space travel have to be acknowledged.
Currently, Mars is within reach for interplanetary habitation based on the current and expected level of technology readiness level likely to be reached in the near future. According to previous research, regolith, crushed rock and dust found on Mars can be potentially used as a construction material (inter al. Spiero and Dunand, 1997; Happel, 1993) and regolith constructions can potentially protect astronauts from large amounts of radiation.

---

[1] http://www.roboticbuilding.eu/project/scalable-porosity/
[2] http://www.kitepower.eu/home.html

However, galactic cosmic rays would require a regolith layer of several meters thick in order to protect the astronauts sufficiently. Furthermore, thermal stresses that occur from large temperature changes during the day night cycle on Mars are a challenge as well as the absence of an atmosphere, which could further increase stresses in the building envelope when creating a pressurized environment.

Precedent case studies counteract these challenges in various ways as for instance, (i) the Mars Ice House developed by the National Aeronautics and Space Administration (NASA) uses ice as main construction material as it is more effective against radiation then regolith-based constructions. Measures are taken to keep the ice from sublimating into the air relying on inflatable plastics[3]; (ii) Foster and Partners autonomous habitation sintering approach uses regolith as main construction material, but instead of printing it they fuse the layers together using an autonomous swarm of robots[4]; (iii) Apis Cor's X-House is a 3D printed habitat that uses Martian concrete reinforced with basalt fibbers and expandable polyethylene foam[5]; furthermore, (iv) AI Space Factory's MARSHA is a 3D printed habitat that used a biopolymer basalt composite material for 3D printing, which is effective against stress and to some degree against radiation as the material has a high hydrogen concentration[6].

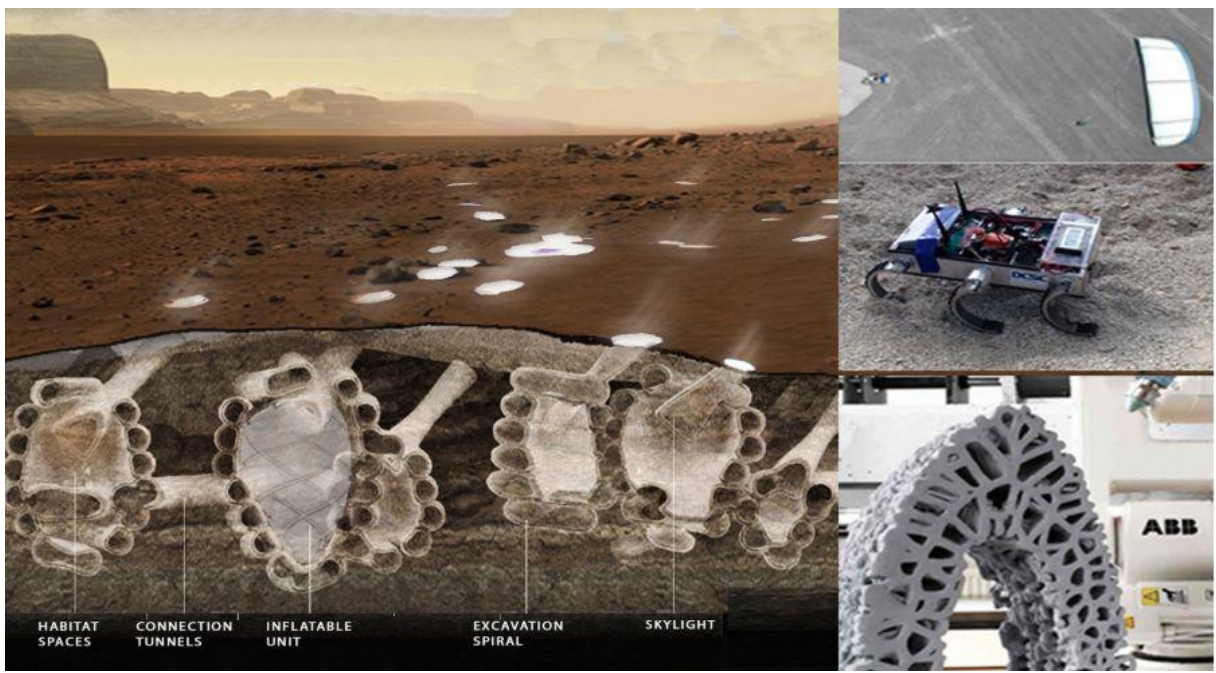

Fig. 1: Underground Martian habitat (left) implemented with D2RP&O methods (bottom right), using rovers (middle right), and relying on renewable energy generation (top right).

---

[3] https://www.nasa.gov/feature/langley/a-new-home-on-mars-nasa-langley-s-icy-concept-for-living-on-the-red-planet
[4] https://www.fosterandpartners.com/projects/lunar-habitation/
[5] http://www.spacexarch.com
[6] https://www.aispacefactory.com/marsha

All these examples are NASA 3D printed habitat proposals and all of them are top surface. The TUD team sees an opportunity to investigate possibilities, how autonomous robots could drill into and/or excavate off-Earth and 3D print/spray a subsurface habitat while also considering the restraints of interplanetary space travel. Main idea is to develop a D2RP method, which facilitates excavation and 3D printing in order to produce subsurface habitats. Subsurface habitation has the advantage of natural protection from radiation while also being less affected by thermal stresses because the temperature is more stable underground. At the same time through excavation valuable resources are mined for in situ resource utilization (ISRU). Design methodologies are, however, restricted by the method of production and the materials available. The use of locally obtained materials through excavation and the naturally obtained shelter represent the advantage over other design proposals.

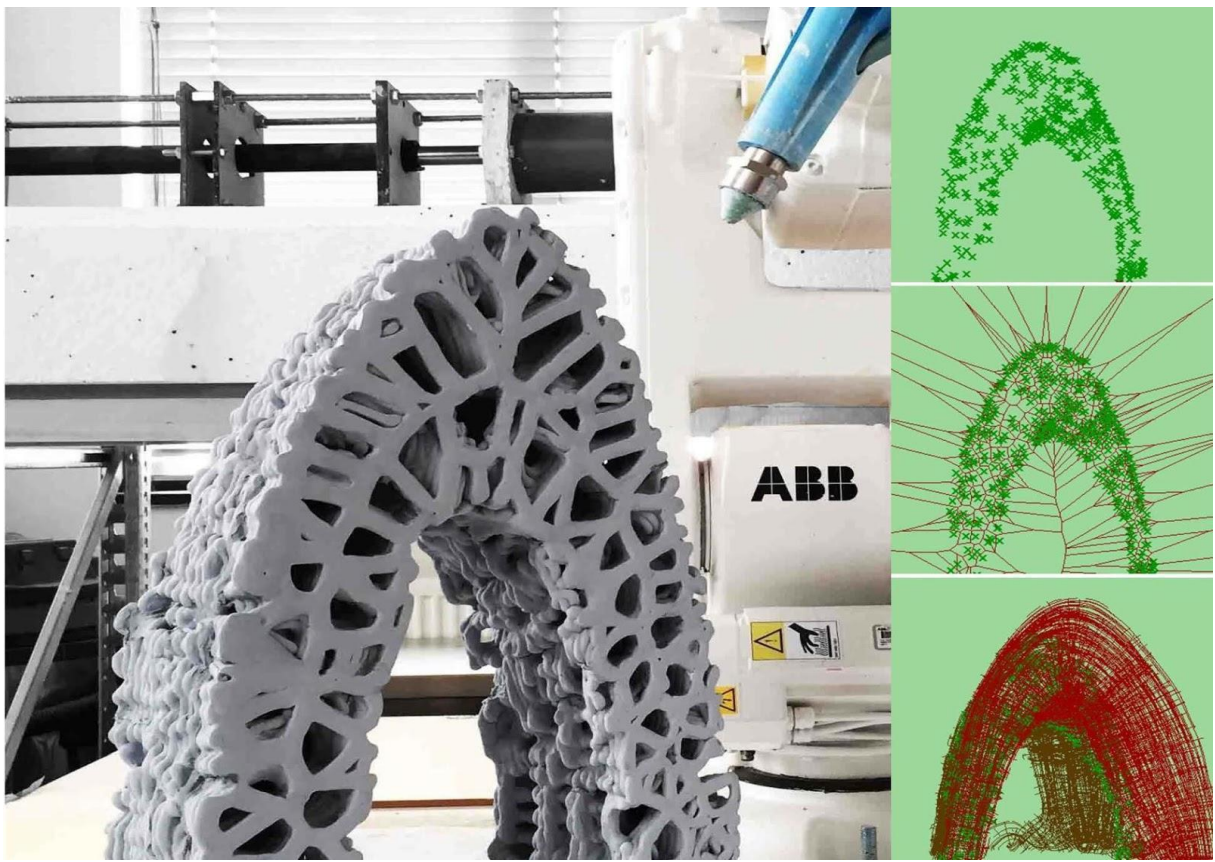

Fig. 2: Additive D2RP using ceramic clay and relying on structural analysis, and robotic path optimisation

The idea developed by the TUD team for the competition Open Space Innovation Campaign 'Off-Earth Manufacturing and Construction', which was put forward by the European Space Agency (ESA), involves autonomous mobile robots that will excavate the ground in a sloped downwards spiral movement. Initially the idea was that the excavated regolith will be mixed with liquid sulphur to create concrete, which is then used by the 3D printing/spraying rovers to stabilize the excavated tunnels. Meanwhile, the team considers using cement-based concrete because of the extensive experience in using this material for building on-Earth habitats as well as the available expertise and technology of the industry partner, Vertico. While cement can be produced on Mars, infrastructure for producing it needs to be already in place leaving the question of, what would be the first structure built from, open. This implies that the structure would be built at a later stage of colonisation with the excavated tunnels first being stabilized using shotcrete in order to, in a second step, remove the material in between the tunnels and create a larger cavity that can be used for inhabitation.

Description

This idea proposes the adaptation of several technologies developed for on-Earth application to off-Earth conditions. This paper focuses on the architecture related aspects and in particular on the Data-driven Design-to-Robotic-Production and -Operation (D2RP&O) processes.

1. Data-driven D2RP integrates advanced computational design with robotic techniques in order to produce architectural formations by directly linking design to building production (Bier et al. 2018). The overall design of the habitat relies on data-driven simulation of the underground rhizomatic structure (figure 1). By analysing the composition of the terrain suitable locations will be identified in order to excavate into the ground in collaboration with experts from Civil Engineering and Geosciences (CEG), TUD. The first case study is for a 60-80 m2 habitat that can be extended in time.

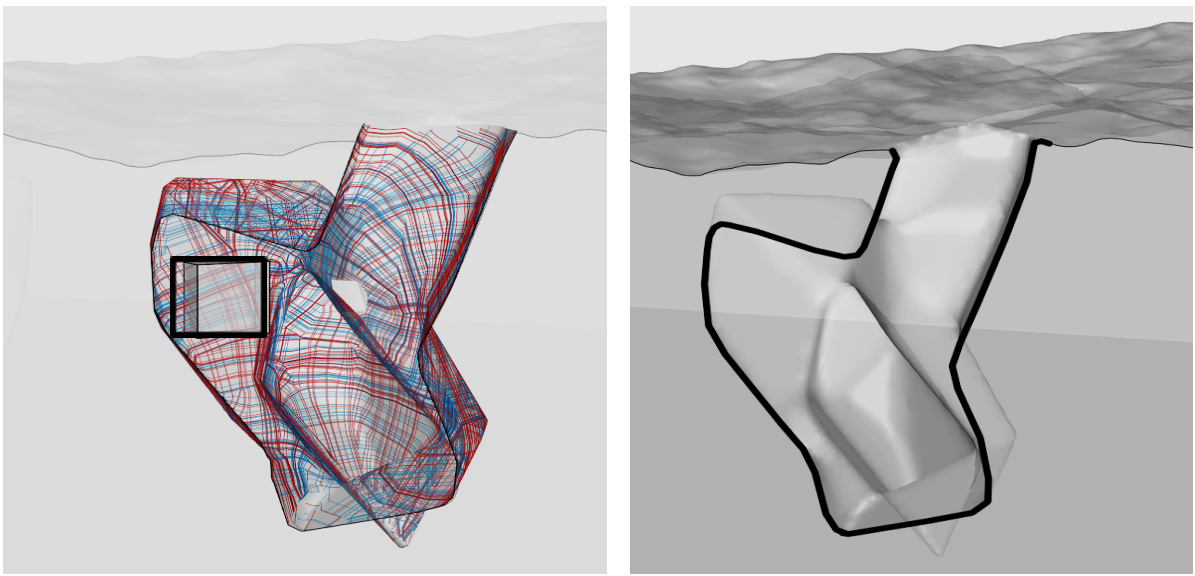

Fig. 3: Section through underground structure (right) and its structural analysis with a fragment chosen for further development (left)

Subtractive and additive D2RP will be employed in the following sequence:

1.1 Subtractive D2RP involving excavation following a regular mining approach developed at CEG is proposed. Excavation is implemented with rovers similar to the ones developed at TUD[7] in a controlled downwards spiral movement (figure 1). D2RP involving milling, drilling, cutting as explored at Robotic Building (RB), TUD[8] will be employed at the later stages of construction when the interior architecture of the habitats is implemented.

---

[7] https://tudelftroboticsinstitute.nl/robots/zebro or
https://www.delftsepost.nl/nieuws/algemeen/772027/tu-delft-studenten-bouwen-aan-eerste-nederlandse-maanrover
[8] http://100ybp.roboticbuilding.eu/index.php/WorkshopHvA

1.2 Additive D2RP as explored in the RB lab with ceramic clay[9] (figure 2) and Thermoplastic Elastomers (TPE)[10] represents the basis for the AM approach with regolith that is proposed in this project. Furthermore, the industrial partner, Vertico, has expertise in robotic 3D printing with concrete. The idea is to connect the printing/spraying system to a swarm of rovers. The assumption is that the generated structure is a structurally optimized porous structure, which has increased insulation properties (figure 2) and requires less material and printing time. Both additive and subtractive D2RP will be powered by the airborne energy[11] system combined with solar cells[12] (Vargas et al., 2021). In order for the built structures to be inhabitable, environmental control, life support and energy requirements need to be considered.

2. Data-driven D2RO links the design to building operation (inter al. Bier et al. 2018). It takes sensor-actuator systems into account that are required for environmental control and life-support. The system supplies air, water and food and relies on filtration systems for human waste disposal and air production requiring an average power for a habitat on Mars of 1600 W for a crew of 6 people (Santovincenzo, 2004). Water needs to be stored, used, and reclaimed (from wastewater), although, Mars missions may also utilise water from the atmosphere or ice deposits. Oxygen comes from electrolysis, which uses electricity from solar panels or kite-power to split water into hydrogen gas and oxygen gas. Temperature regulation is achieved using both passive and active systems, which protect from overheating, either by thermal insulation and by heat removal from internal sources (such as the heat emitted by the internal electronic equipment) or protect from cold, by thermal insulation and by heat release from internal sources. Furthermore, shielding against harmful external influences such as radiation and micro-meteorites is necessary. This is achieved by placing the habitat below ground level. In addition, an inflatable structure is proposed to counteract Mars's low atmospheric pressure, which is a threat to human health. The inflatable structure that regulates the indoor pressurised environment is placed in the 3D printed structure. This inflatable structure requires materials such as neoprene, vectran, kevlar, etc. These materials can also be reproduced on Mars through ISRU of silicon, which is proven to be available in abundance on Mars.

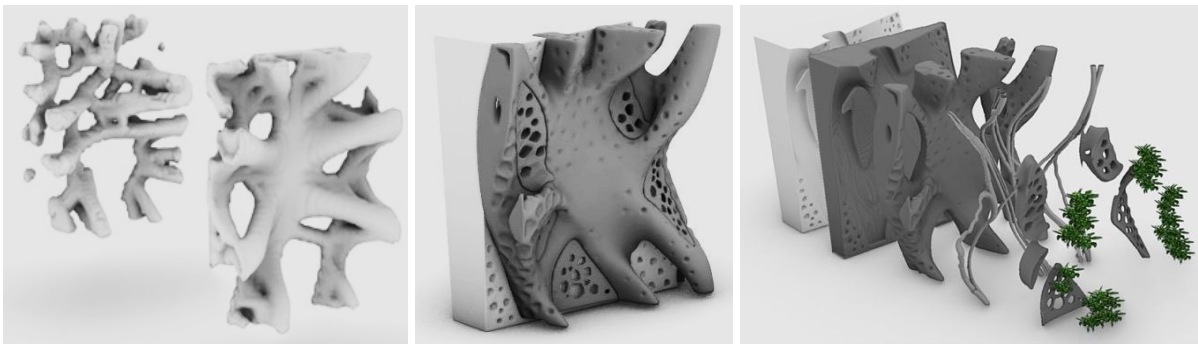

Fig. 4: Structure based on stress lines analysis (left), structural mesh with acoustic optimization pattern (middle), and exploded axonometric view showing excavated layer, 3D

---

[9] http://www.roboticbuilding.eu/project/scalable-porosity/
[10] http://www.roboticbuilding.eu/project/variable-stiffness/
[11] http://www.roboticbuilding.eu/project/kite-powered-d2rp/
[12] https://www.tudelft.nl/en/eemcs/the-faculty/departments/electrical-sustainable-energy/photovoltaic-materials-and-devices/

printed/sprayed layer, sensor-actuators and plant pods layers (right)

Life support systems could include a plant cultivation system, which could also regenerate water and oxygen. Such a system could reuse nutrients via composting waste which is then used to fertilize crops. Research results of the Micro-Ecological Life Support System Alternative (MELiSSA)[13], which is a ESA led initiative, aiming to understand the behaviour of artificial ecosystems in order to develop technology for a future regenerative life support system, could be integrated in the proposed project. The life-support system is, however, not discussed in this paper in more detail.

Energy generation by means of solar panels and airborne wind energy is considered as described by Vargas, et al. (2021) and is not discussed in this paper. Instead the plant cultivation system is shortly addressed.

Case study
The proposed idea has been tested in a case study by means of the D2RP&O, which involved the development of a 1:1 scale prototype as a proof of concept for habitat envelope components that facilitate cultivation of plants. The structure is analysed with respect to structural stresses and the output result is derived in the form of tension and compression lines (figure 3). A 1-meter by 1-meter fragment of the envelope is selected for further development. The fragment is developed considering structural loads and the result is topologically optimized (figure 4). This structural mesh is designed based on the analysis result and the integration of sensor-actuators that enable and regulate plant growth.
The underground is excavated and is reinforced with concrete. The milled and reinforced surface serves a dual purpose, it supports the cave-like structure and also has acoustical properties generated through surface tectonics (figures 4 and 5). The structural mesh is sprayed and/or 3D printed. The sensors and wires are integrated, while detachable 'plant pods' for hosting plants are overlaid on the structural mesh (figure 5). The detachability is for the easy maintenance. A selection of plants such as lettuce, basil, mint, dill, rosemary, thyme, and soybeans was identified in dialog with the University of Wageningen (Wamelink et al., 2014).

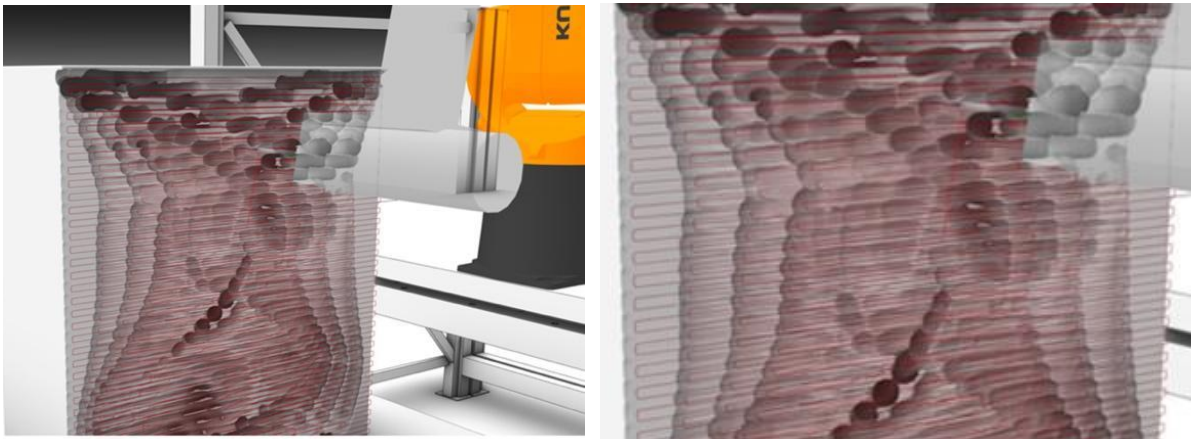

Fig. 5: Simulated toolpaths for subtractive D2RP.

---
[13] https://www.esa.int/Enabling_Support/Space_Engineering_Technology/MELiSSA_life_support_project_an_innovation_network_in_support_to_space_exploration

The robotic production was digitally simulated to optimize the robotic path and the first tests were implemented by milling expanded polystyrene (EPS) and 3D printing with fiber reinforced TPE. The aim was to emulate the process that is to be implemented on Mars. The tests identified the requirement to adjust surface tectonic because the material deposition during printing introduced a considerable flattening of the acoustic patterns. In the next step, prototyping will be implemented with improved patterns and with commercially available regolith simulants and cement. The 3D printed porous structure will be then compared with a non-porous structure in order to identify differences in structural and insulations properties.

Conclusion:
The presented project investigates possibilities to develop Maritan subsurface habitats while also considering the restraints of interplanetary space travel. The paper presents the overall framework within which such an approach could be implemented and discusses a case study that describes the subtractive and additive D2RP approach. While digital prototyping has been successfully implemented, the physical robotic prototyping is still work in progress.

Acknowledgements:
This paper has profited from the contribution of teams from the faculties of Aerospace Engineering, Maritime, Material, and Mechanical Engineering, Civil Engineering, and Architecture. Furthermore, it has profited from the contribution of Wieger Wamelink from the University of Wageningen and Advenit Makaya from ESA.